# Experimental Observation of Energy Modulation in Electron Beams Passing Through Terahertz Dielectric Wakefield Structures

S. Antipov[1,3], C. Jing[1,3], M. Fedurin[2], W. Gai[3], A. Kanareykin[1], K. Kusche[2], P. Schoessow[1], V. Yakimenko[2], A. Zholents[4]

[1]. Euclid Techlabs LLC, Solon, OH 44139

[2]. Accelerator Test Facility, Brookhaven National Laboratory, Upton, NY 11973

[3]. High Energy Physics Division, Argonne National Laboratory, Lemont, IL 60439

[4]. Advanced Photon Source, Argonne National Laboratory, Lemont, IL 60439

Abstract:

We report observation of a strong wakefield induced energy modulation in an energy-chirped electron bunch passing through a dielectric-lined waveguide. This modulation can be effectively converted into a spatial modulation forming micro-bunches with a periodicity of 0.5 – 1 picosecond, hence capable of driving coherent THz radiation. The experimental results agree well with theoretical predictions.

Free electron laser (FEL) based Terahertz (THz) source technology is considerably attractive because of its capabilities for producing high peak power (beyond MW) at a high repetition rate (beyond MHz) [1]. The key to generating coherent radiation in THz FELs is the formation of sub picosecond micro-bunches that are used to drive the THz radiation source. In the past decade, many approaches have been investigated to generate THz micro bunches that include: bunch generation from a photoinjector with micro laser pulses produced by birefringent crystals [2]; bunch train with a picosecond separation using an emittance exchanger combined with transverse beam masking and other similar techniques [3, 4]; and some bunch compression techniques [5, 6].

In this paper, we report on the successful results of producing a strong energy modulation of an electron beam by means of the self-wake excited in a simple dielectric-lined waveguide. We used cylindrical geometry dielectric wakefield structures in this set of experiments. Alternatively, other geometries can be used, for example rectangular / planar. Planar geometries with adjustable beam gaps for tuning the wakefield spectrum [7] can be used to produce tunable energy modulation. The energy modulation can be further transferred to density modulation by passing the beam through a chicane which is normally used for pulse compression of energy chirped beams [for example 6]. In this experiment the energy modulation is produced by a THz structure, hence sub-picosecond bunch trains can be produced out of this beam utilizing only dipole magnets without further compression. The density modulated beam (a bunch train) can later be used as a drive beam for wakefield acceleration in THz structures [8] or for coherent emission of radiation [6, 9, 10]. It can be further compressed for applications in FELs and plasma wakefield acceleration [4, 6, 12]. This approach fills the niche between microbunching (a periodicity of a few microns) by inverse FEL acceleration [12, 13, 14] and bunching by laser pulse stacking for photoinjectors (mm periodicity) [2]. Similar microbunching can be achieved by masking the energy-chirped beam which travels through the dogleg. The mask is placed between the dipoles, in the region where the beam transverse size is dominated by the correlated energy spread [4]. This method has the disadvantage of partial beam loss at the mask.

The principle of introducing an energy modulation in the beam is rather simple. The self-wake in general reduces the beam quality: the transverse wake deflects the beam and the longitudinal wake introduces an energy spread. However, when the bunch length is comparable or much longer than the wavelength of the fundamental mode of the wakefield, the wakefield inside the bunch will show an amplitude modulation, particularly for a triangular or rectangular shaped bunch [15].

Intrinsically, without further corrections the electron bunch develops an energy chirp when it is accelerated in linacs. This energy chirp helps establish the energy bands around every amplitude cycle of the wakefield inside the bunch. Fig. 1 shows an example of the theoretical process to form the energy modulation for a positive chirp (higher energy at the tail) triangular bunch. It should be pointed out that the conditions we imposed in this example are not strict at all. The energy chirp can have either positive or negative slope; and the bunch shape can be other shapes as well, like a long rectangular bunch which can be easily generated using the laser stacking technique [2].

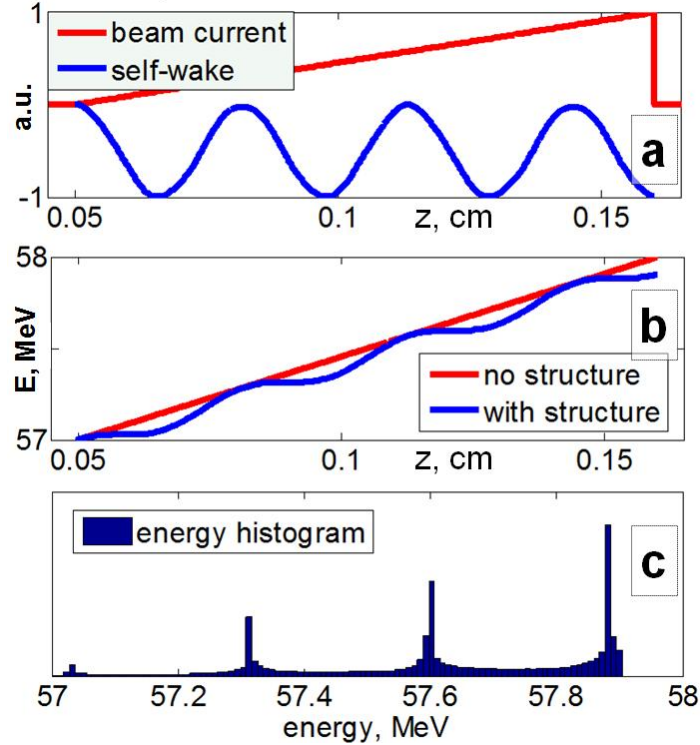

FIG. 1. a) Beam current profile and self- induced wake inside the beam. b) Energy – longitudinal coordinate distribution; original (red) and modified due to self-deceleration (blue). c) Histogram of the self-energy-modulated beam.

As shown in Fig. 1(a), a triangular shaped bunch has a current density distribution

$$I(t) = I_0 \omega_0 t \text{ for } 0<t<T \text{ and } I(t)=0 \text{ otherwise.} \qquad (1)$$

Here $I_0$ is a constant, and $\omega_0$ is the frequency of the lowest mode of the wakefield device. For simplicity, we consider a single mode wakefield structure for this argument, while the numerical model [Fig. 1] takes in account first 8 higher order modes. The self-wake inside the drive bunch is [16]

$$V^-(t) = -\int_0^t 2kI_0\omega_0\tau \cos[\omega_0(t-\tau)]d\tau = -\frac{2kI_0}{\omega_0}(\cos\omega_0 t - 1) \quad (2)$$

which shows that the particles lose different energies in the wakefield device depending on their position in the bunch. Particles at nodes, $t=\pi N/2\omega_0$ do not lose energy at all. If the bunch has a linear energy chirp at entrance of the wakefield device, its longitudinal phase space will form a staircase shape [shown in Fig. 1(b)]: a strong energy modulation appears in the spectrometer after the device [Fig. 1(c)], but its temporal profile is preserved for a relativistic beam.

The experiment was performed at the Accelerator Test facility (ATF) at Brookhaven National Laboratory, which can provide an adjustable length shaped bunch with a linear energy chirp. For this experimental setup a 130 pC beam with about 1.6 mm length was used. The beam energy is 57 MeV with 1 MeV energy chirp. The beam current is shaped by means of a mask inserted in a region of the beam line where the beam transverse size is dominated by the correlated energy spread [4, 9]. The beam shaping mask was made in a form of an arrow: a triangular hole followed by a rectangular channel, which was originally designed for the requirements of another experiment [8]. In this experiment we were able to partially block the mask, producing beams ranging from 250 micron to 800 micron long triangular pulses and a 1.6 micron arrow shape pulse. These dimensions were measured and calibrated by coherent transition radiation (CTR) interferometry. Wavelengths longer than the beams are emitted coherently and carry information about the bunch lengths [17, 18]. Transition radiation is sent to the interferometer and the signal is recorded by a helium-cooled bolometer [4, 9]. Because of the way the beam is shaped there is a linear energy chirp from the head of the beam to its tail. In the energy dispersion – free beamline optics this beam can be transported downstream to the spectrometer and maintain its "arrow" shape on the spectrometer screen [Fig. 2(a)].

In this experiment we used three different quartz capillary tubes as wakefield structures. Each was metallized via gold sputtering on the outer surface and inserted into a stainless steel tube and then into a motorized holder. The dielectric constant of quartz is 3.8 over a broad range of frequencies including the THz range [19]. The dimensions and their synchronous $TM_{01}$ mode frequencies (inner diameter, μm / outer diameter, μm / frequency, THz / length, mm) are: 1) 330 / 390 / 0.95 / 25.4; 2): 230 / 330 / 0.76 / 25.4; and 3): 450 / 550 / 0.615 / 51. The dimensions were measured using a microscope. The thicknesses of the quartz tubes are small enough to diminish the effects of all high order modes.

The total "arrow" beam was 1.6 mm (0.8 mm – triangle part and 0.8 mm rectangular part) long; much longer than wavelength of three wakefield structures. The self-induced wake inside the "arrow" beam can modulate the energy producing as many "energy bunchlets" as there are wavelengths per total beam length. In Fig. 2, we observe energy modulation into 6 bunchlets using the 0.95 THz structure, 5 for 0.76 THz and 4 for 0.615 THz. The features in the last case are particularly interesting: the first two "energy bunchlets" are split in two. This feature will be explained by a simpler measurement of a shorter triangular bunch.

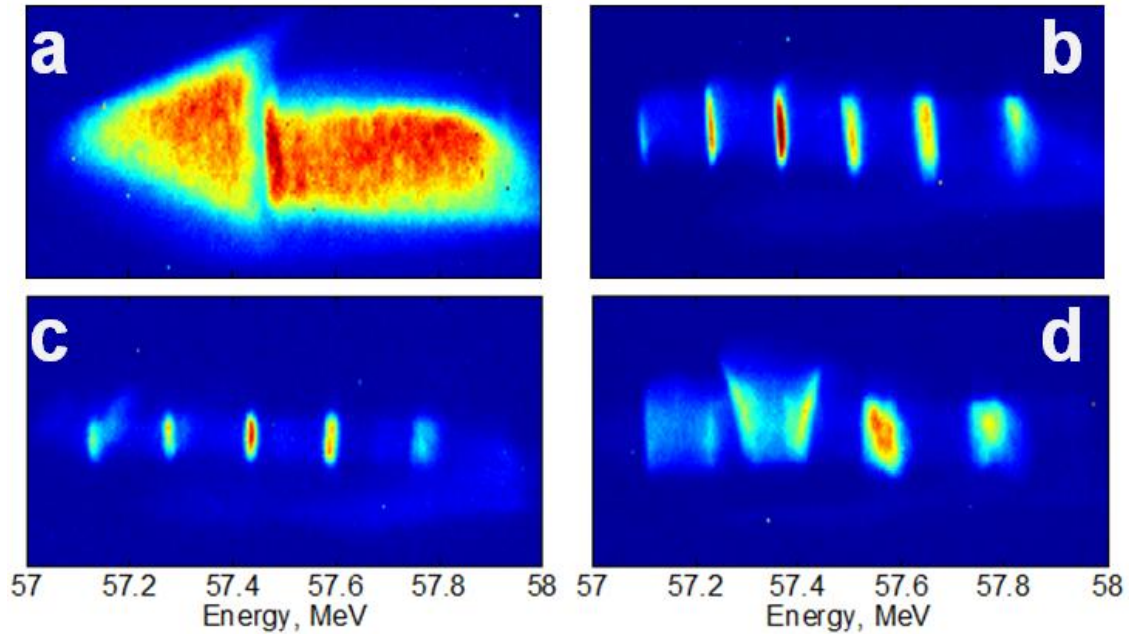

FIG. 2. Spectrometer images of the full-size 1.6mm long "arrow" beam (triangle followed by rectangle). a) original, undisturbed beam. b) Beam passing through 0.95 THz structure. c) Beam passing through 0.76 THz structure. d) beam passing through 0.615 THz structure.

If the length of the structure is longer than needed then trailing particles in a bunch may be decelerated to a lower energy than the leading particle (for which the energy stays the same). In this case we observe double-bunches [Fig. 3(b)]. We observed this in the case of a 65 pC total charge, 800 micron triangular beam passing through a 25.4 mm long, 0.95 THz wakefield structure. First of all, the beam is about 3 wavelengths in size; hence we see energy modulation into three energy bunchlets. Another way to observe it is to note on Fig. 3(c) that there are three locations in the triangular beam which don't experience the self-wakefield (the head of the triangle, $z \approx 0.05$ cm, $z \approx 0.082$ cm and $z \approx 0.114$ cm). Because energies of particles at these locations are different they will produce three energy bunchlets when other particles around them start experiencing deceleration and cluster around them. This example also shows the importance of the initial energy chirp for production of multiple energy bunchlets. If there were no chirp, particles at z ≈ 0.05 cm, z ≈ 0.082 cm and z ≈ 0.114 cm would have the same energy and location on the spectrometer. The observed three bunches will merge into one if the energy chirp is not present.

Comparing Fig. 1(b) to Fig. 3(c) the structure is longer than optimal for energy bunching (wakefield deceleration is too strong). This leads to the appearance of double bunches, see Fig. 3 (b) – experiment, (d) - simulation. Spectrometer resolution (estimated from the smallest energy spectrum image features at about 65 keV – full width at half maximum, FWHM) prevents us from observing sharp energy peaks, like on figure 3(d).

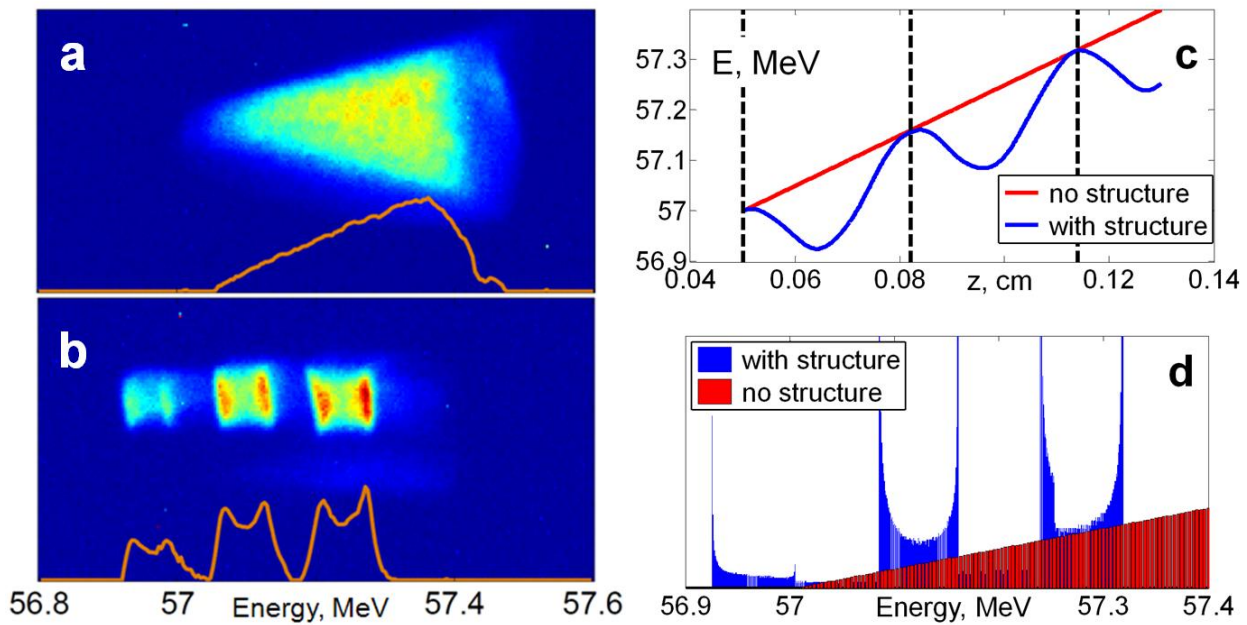

FIG. 3. a) spectrometer image and its projection of an unperturbed, 800 micron long beam. b) spectrometer image and its projection of the same beam that passed through the structure (measurement). Energy modulation is observed. c) Simulated energy – longitudinal coordinate distribution; original (red) and modified due to self-deceleration (blue). d) Simulated histogram of original beam (red) and self-energy-modulated (blue).

In the simulation [Fig. 3(c)] we observe that particles, immediately following the leading particle are decelerated further below 57 MeV. Particles at z ≈ 0.082 [Fig. 3(c)] don't experience deceleration; hence they will form the second energy bunchlet. Particles immediately following them get decelerated to have energy less than the particles at z ≈ 0.082. This forms the double bunching observed in the experiment [Fig. 3(b)]. Double bunching may not be an effective way to produce a modulated beam because of the loss after the chicane of particles that are not bunched. However, this allows for two bunches per wavelength for a particular structure, hence higher frequency bunching. In this case it is 6 bunches per 800 micron beam obtained in a 0.95 THz structure. When transferring this energy modulation into a density modulation, the chicane can correct or enhance some energy modulation features obtained using the wakefield device. For example three *double* bunchlets in the energy spectrum observed experimentally due to excessive deceleration in the wakefield device [Fig. 3(b)] can be transferred into three *single* bunches in density when using a chicane [20].

Finally, in beams smaller than a wavelength only a single energy bunchlet is produced. Hence this approach can be used for reducing energy spread – a concept dubbed the "wakefield silencer" at the Argonne Wakefield Accelerator Facility [15]. At ATF we used a 247 micron triangular beam with a 200 keV energy spread (FWHM) which was reduced by almost a factor of three to 70 keV by passing it through the 0.95 THz structure [Fig. 4]. In principle, by choosing a smaller size triangle one can reduce the energy spread even further. However the spectrometer resolution limit prevents us from observing a much narrower energy spread in this experiment.

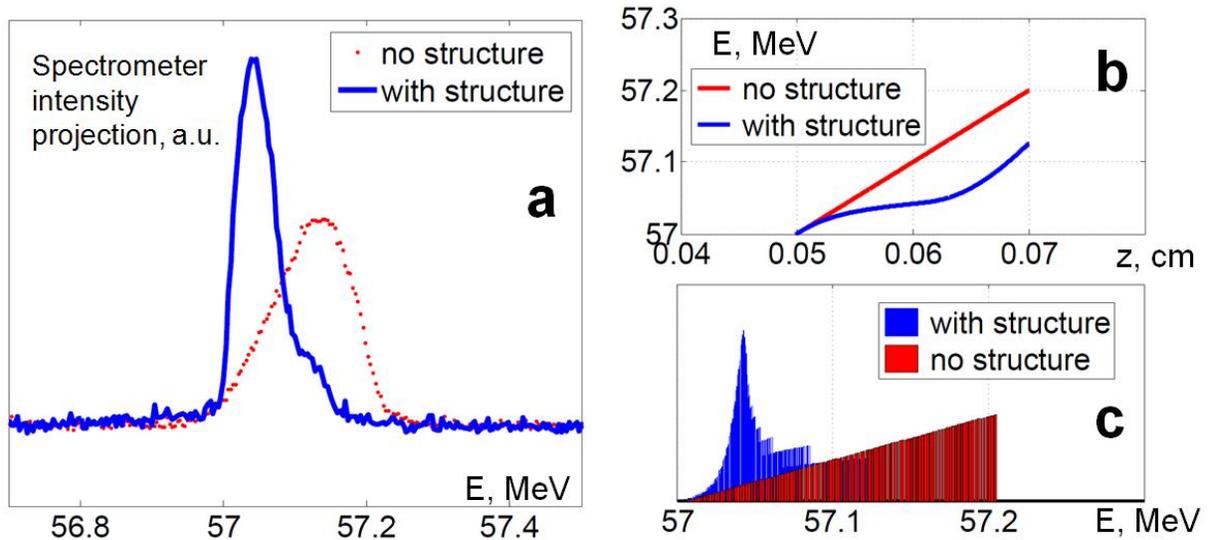

FIG. 4. a) Spectrometer intensity projection (measurement) of original (red, dotted) and modified due to self-deceleration (blue solid) beam. b) Simulated energy – longitudinal coordinate distribution; original linear chirp (red) and compressed due to self-deceleration (blue). c) Simulated histogram of original beam (red) and self-compressed (blue).

For effective energy modulation the structure length has to be chosen appropriately with respect to the beam charge (self-wake gradient) and the energy chirp to produce the case shown in Fig. 1(b, c), rather than Fig. 3(c, d). Identically, the total charge of the beam, with all the dimensions staying the same, can be adjusted for a particular structure length. This approach is more convenient in the actual experiment. The combination of beam charge, structure length and initial energy spread can be packaged in a quantity that we term the "silencer strength" for the case of short beam's energy spread reduction. We illustrate it in the following example. Wakefield silencing (reducing the energy spread by means of passive wakefield structure) can be utilized to reduce the energy spread of the FACET beam [21], which is almost 1 GeV. Considering the small dimensions of the FACET beam ($\sigma_z$ = 30 μm) any E-201 collaboration structure proposed for FACET studies [22] can be used as a silencer. Assuming that the FACET beam can be shaped to be triangular beam, 30 micron long with 2 nC total charge we will employ a multimode alumina structure with 508 micron ID and 790 micron OD. Such a beam will produce an almost linear (similar in shape to the area from z = 0.05 to z = 0.055 on Fig. 1 (a) decelerating field inside the triangle peaking at the tail with about 800 MV/m. This means that original 1 GeV energy spread can be shrunk to 750 MeV using a 25 cm long silencer. If the silencer strength is doubled (by length or total charge) the energy spread will be reduced to 500 MeV.

In the measurements reported here we worked mostly with triangular beams. Energy modulation similar to the ones reported in this paper can be produced in beams with other current distributions, like rectangular bunch. For example, Fig. 2 shows, that energy modulation was observed also in the rectangular part of the "arrow" beam.

In summary, this paper demonstrated that an energy chirped beam can experience a strong energy modulation by self-wake when passing through a passive wakefield structure. The number of energy bunchlets depends on the frequency of the wakefield structure, its length and beam's energy chirp and charge. Our numerical model accurately explains results including double bunching for the case when the wakefield structure length is not optimal for the beam. A beam with an initially flat energy distribution can only be modulated with two energy peaks (double bunching) with modulation depth defined by the length of the structure and beam total charge. The energy modulation observed during the experiment can be effectively converted into a spatial modulation forming micro-bunches with a periodicity of 0.5 – 1 picosecond, capable of driving coherent THz radiation. This conversion can be done by chicane allowing for additional tuning and correction of micro bunching. Using a passive wakefield device together with a chicane is a simple and effective way of producing micro-bunched beams for beam based high power THz sources. Utilization of tunable wakefield structures for both energy modulation and radiation allows for adjustable micro-bunching hence tunable THz sources.

Acknowledgements: Euclid Techlabs LLC acknowledges support from DOE SBIR program grant #DE-SC0006299.